\documentclass[
    ,final            
  ]
  {aipproc}

\layoutstyle{6x9}

\begin{document}
\title{QCD VACUUM AND CONFINEMENT}
\author{Adriano Di Giacomo}{
  address={University of PISA and INFN}
}

\begin{abstract}
This course consists of two lectures. In the first lecture I
discuss why a non perturbative formulation of QCD is needed ,and I 
show
that lattice formulation copes with this need, even if it mainly 
produces
numerical results. In the second lecture I discuss how lattice can 
help
to understand the deconfinement transition.Such understanding is also
important to predict parameters that can help in the interpertation of
heavy ions high energy experiments.
\end{abstract}
\maketitle
\section{Introduction}
Vacuum is by definition the ground state of a field system:it is 
stable
against quantum fluctuations. 

An exact knowledge of the ground state
provides complete information on the system.This property goes under 
the
name of reconstruction theorem,the exact statement being that from the
field correlators
\begin{equation}
\langle0| T(\varphi_1(x_1)\ldots\varphi_n(x_n)|0\rangle
\label{eq1}\end{equation}
the Hilbert space and the matrix elements of physical observables can 
be
constructed.

Textbook quantization is perturbative .The Lagrangean is split in a 
free
term  ${\cal L}_0$   and a perturbation ${\cal L}_I$
\begin{equation}
{\cal L} = {\cal L}_0 + {\cal L}_I
\label{eq2}\end{equation}
 ${\cal L}_0$ is exactly solvable ,and defines as vacuum the ground 
state of Fock
space,i.e. empty space.    ${\cal L}_I$   is a small perturbation 
producing
scattering between fundamental particles,
and small changes in the 
wave
function of the ground state. 

In QCD ${\cal L}_0$       describes 
free gluons and
quarks and  their interactions. For sure this is not a good starting 
point
,since quarks and gluons are confined.Fock vacuum is not the ground
state,and therefore it is not stable against perturbations.

This is most
probably the reason why the perturbative expansion is not 
convergent,not
even as an asymptotic series,in spite of the fact that it seems to 
work
at small distances\cite{1}.

The knowledge of the true vacuum is,in particular,relevant to the
understanding of one of the most intriguing properties of QCD, namely
confinement of color. 

In spite of the well established evidence that
quarks and gluons are the fundamental constituents of hadronic matter,
they have never been observed in Nature as free particles. 

All 
particles
produced in particle reactions are color singlets.This property is 
known
as confinement of color.

Since the pioneering papers of Gellmann in which quarks were 
introduced
as fundamental constituents of hadrons ,quarks have been searched for  in
particle reactions and in Nature.None has been found,so that 
experiments
establish upper limits,which can be found in  Particle Data Group\cite{2}.
The cross section   $\sigma_q$  for the inclusive reaction
\begin{equation}
p + p \to q\,(\bar q) + X\label{eq3}\end{equation}
  in which a  $q,\,(\bar q)$ is detected as a fractionally charged 
particle has an
upper limit
\begin{equation}
\sigma_q < 10^{-40}\,{\rm cm}^2\label{eq4}\end{equation}
to be compared to the total cross section  $\sigma_T$   at the same 
energy      
$\sigma_T \sim 10^{-25}\,{\rm cm}^2$.
In the absence of confinement $\sigma_q/\sigma_T$          would be 
of the order of
unity,and is instead   $\sigma_q/\sigma_T\simeq 10^{-15}$.

Similarily the abundance $n_q$   of free quarks in Nature is measured 
by
looking for particles of fractional charge    $q = \pm1/3, 
\pm2/3$      in Millikan
like experiments.

No quarks have been found,and the resulting upper 
limit
is     $n_q/n_p < 10^{-27}$   corresponding to the analysis of    
$\sim   1g$ of
matter. 

In the absence of confinement the Standard Cosmological Model
(SCM) predicts   $n_q/n_p \sim 10^{-12}$\cite{3}. Again a factor  
$10^{-15}$       
 between upper limit and
expectation. 

Such a small factor cannot have a natural theoretical
explanation ,except if the actual value of the above quantities is
exactly zero.Confinement is an absolute property,and this can only be
explained in terms of a symmetry property of the vacuum.The knowledge 
of
the true vacuum is necessary to understand confinement .A quantization
procedure that is not based on perturbation theory is needed.

\section{Feynman path integral\cite{4} and lattice formulation of QCD\cite{5}.}
A quantum system is defined by the canonical variables ,$q ,p$ and the
Hamiltonian $H$. Solving the system means to construct a Hilbert 
Space on
which  $q ,p$ act as operators obeying the equations of motion and the
canonical equal-time commutation relations.
\begin{equation}
[q_i,p_j] = i \delta_{ij}\label{eq5}\end{equation}
A ground state must exist.

Field theory is a special case. The fields $\varphi_i(\vec 
x,t)$         play the role of the
$q$'s ,their conjugate momenta   $\Pi_i(\vec x,t) = \partial{\cal 
L}/\partial
(\partial_0\varphi_i(\vec x,t))$             the role of  the $p$'s.
The system has infinitely many degrees of freedom.To simplify the
notation we shall refer in what follows to a system with one  $q$  
and one
$p$ ,since our arguments will apply to any number of degrees of
freedom. Problems with divergencies ,arising from infinite number  of
degrees of freedom are technical in nature and do not affect the 
arguments
below.The same is for problems in defining the conjugate momenta,
which
are a consequence of gauge invariance.

Since the  $|q\rangle$'s are a complete set of states ,the knowledge of 
the amplitudes
\begin{equation}
A(q',t',q,t) = \langle q'|e^{-i 
H(t'-t)}|q\rangle\label{eq6}\end{equation}
contains all physical information.

Let us divide the time interval  $t'-t$  into $n+1$ intervals of 
equal length $\delta$,
\[ \delta = \frac{t'-t}{n+1}\]
$\delta$  is a small quantity in the limit of large  $n$. 
We can write
\begin{equation}
e^{-i H(t'-t)} = e^{-i H\delta}e^{-i H\delta}\ldots e^{-i H\delta}
\label{eq7}\end{equation}
the product of $n+1$ factors.

Again for the sake of notational simplicity we shall assume that
\begin{equation}
H = \frac{p^2}{2m} + V(q)
\label{eq8}\end{equation}
The generic case with linear terms in $p$ ,or $q$-dependent 
coefficients of $p^2$,
$p$   only present technical complications ,but do not alter the 
results
which follow. 

The amplitude eq (6) becomes, by inserting eq (7) and a
number of projectors on a complete set $\int dq |q\rangle\langle q| = 
1$
\begin{equation}
A = \int dq_1\ldots dq_n\langle q'|
e^{-i H\delta}|q_n\rangle\langle q_n | e^{-i H\delta}|q_{n-1}\rangle
\ldots \langle q_1 | e^{-i H\delta}|q\rangle
\label{eq9}\end{equation}
On the other hand ,by use of the well known Baker-Haussdorf formula
\begin{equation}
e^{-i H\delta}
\simeq e^{-i\frac{p^2}{2m}\delta} e^{-i V(q)\delta}\left[ 1 + {\cal 
O}(\delta^2)\right]
\label{eq10}\end{equation}
so that 
\[
\langle q_{i+1} | e^{-i H\delta}| q_i\rangle =
e^{-i V(q_i)\delta} \langle q_{i+1} | e^{-i\frac{p^2\delta}{2m}}| 
q_i\rangle
\left[1+{\cal O}(\delta^2)\right]
 \]
and since
\[
 \langle q_{i+1} | e^{-i\frac{p^2\delta}{2m}}| q_i\rangle =
\int\frac{dp}{2\pi}  \langle q_{i+1} | p\rangle \langle p|q_i\rangle
e^{-i\frac{p^2\delta}{2m}} =
\int\frac{dp}{2\pi} e^{ip(q_{i+1}-q_i) - i\frac{p^2\delta}{2m}} =
e^{i \frac{m}{2\delta}(q_{i+1}-q_i)^2}\]
finally
\begin{equation}
\langle q_{i+1} | e^{-i H\delta}| q_i\rangle \simeq
e^{i\left[\frac{m}{2}\frac{(q_{i+1}-q_i)^2}{\delta^2} - 
V(q_i)\right]\delta}
= e^{i\delta\,{\cal L}[q_i]}
  \label{eq11}\end{equation}
Inserting eq(11)   into eq (9)
\begin{equation}
A = \lim_{n\to \infty}\int\left[\prod_{1}^n d q_i\right]
e^{i\int_t^{t'}{\cal L} dt}
  \label{eq12}\end{equation}
In the limit $n\to\infty$, $\delta\to 0$           and eq(11) becomes 
exact. Eq(12) is usually
written  as
\begin{equation}
A = \int\left[{\cal D} q\right]
e^{i S[q',t',q,t]}
  \label{eq13}\end{equation}
The limit in eq(12) defines an integral over an infinite number of
variables (a functional integral) if it exists.The amplitude eq(6) is 
an
integral over all paths from $(q,t)$  to $(q',t')$  and is defined as 
a
limit of a lattice in time ,as the points of the lattice fill densely
the time interval $(t,t')$.

Exponentials in eq(11) are analytic functions : the construction 
allows
an analytic  continuation of the amplitude to euclidean time  $-t \to 
\tau = -i t$
If $J(t)$ is an arbitrary function in the time interval $(t,t')$ we 
define
\[S_J \equiv S(q,t,t') + \int_t^{t'} dt q(t) J(t)\]
\begin{equation}
A_J(q',t',q,t) =
\int[{\cal D}q] \, \exp\left[i S_J(q',t',q,t)\right]
 \label{eq14}\end{equation}
In the discretized version,
\[ \int_t^{t'} dt q(t) J(t) = \sum_{i=1}^n q(t_i) J(t_i)\]
We define the amplitude
\begin{equation}
A(t',t_1\ldots t_k, t) =
\langle q',t'| T(q(t_1)\ldots q(t_k))|q,t\rangle\qquad t \leq t_i\leq 
t'
  \label{eq15}\end{equation}
by the same construction which led to eq(13) as a limit of a 
discretized
integral
\[A(t',t_1\ldots t_k, t) =
\int [{\cal D}q] \,q(t_1)\ldots q(t_k) e^{-i S[t',t]}\]
It follows from the definition of $A_J$, eq.(14), that
\begin{equation}
A(t',t_1\ldots t_k, t) =
i^n \frac{\delta^n}{\delta J(t_1)\ldots \delta J(t_k)}
\left.A_J(q',t',q,t)\right|_{J=0}
  \label{eq16}\end{equation}
We shall now compute the amplitude for immaginary 
times                 
$A_J(q',-iT; q,i T)$,
in the limit $T\to\infty$ . We shall assume that the external 
current  $J$  is
different from zero only in the time interval                   .
$-T/2 \leq t \leq T/2$.
Then, calling    $|E_n\rangle$   a complete set of states with 
definite energy,
\begin{equation}
A_J(q',-iT; q,i T) =
\sum_{E_n,E_{n'}}
\langle q'|e^{-HT/2}|E_n\rangle\langle E_n 
-i\frac{T}{2}|E_{n'},i\frac{T}{2}\rangle_J
\langle E_{n'}| e^{- HT/2}|q\rangle
  \label{eq17}\end{equation}
since by assumption from $-T$       to  $-T/2$  and from  $T/2$  to  
$T$      the evolution
is governed by $H$.  In the limit of large  $T$
\begin{equation}
A_J(q',-iT; q,i T) \mathop\simeq_{T\to\infty}
e^{-E_0T} \psi_0(q') \psi_0^*(q) \langle 0|
\int[{\cal D}q]\, e^{-S[J]}|0\rangle
  \label{eq18}\end{equation}
$|0\rangle$   is the ground state and $E_0$   its energy. Excited 
states are exponentially
depressed in the sum eq(17).Modulo a multiplicative constant which 
will
be shown to be unimportant,the amplitude of eq(16) is then equal to
\begin{equation}
Z[J] = \langle 0|
\int [{\cal D}q]\, e^{-S[J]}|0\rangle
  \label{eq19}\end{equation}
  and,as in eq's(15),(16)
\begin{equation}
\langle 0| T(q(t_1)\ldots q(t_k))|0\rangle =
\left. \frac{\delta^n Z[J]}{\delta J(t_1)\ldots\delta 
J(t_k)}\right|_{J=0}
  \label{eq20}\end{equation}
$Z[J]$ is known as the functional generator of the field correlators
(20).

Since the field correlators fully describe the theory,knowledge 
of
$Z[J]$ means solution of the system.In particular the euclidean 
Feynman
integral uniquely identifies the true ground state of the 
theory.

Feynman
integral quantization is more fundamental than perturbative 
quantization.

For QCD the euclidean functional integral will be
\begin{equation}
Z[J] = \int[{\cal D} A_\mu] [{\cal D}\psi] [{\cal D}\bar\psi]
e^{-S_J[A,\bar\psi,\psi]}
  \label{eq21}\end{equation}
where the role of  $q$ is plaied by the gluon fields  $A_\mu$  ,the 
quark field $\psi$
and its hermitian conjugate $\bar\psi$.  The integral is defined by 
discretizing
a finite volume  $V$  of space time to a lattice of spacing  $a$  and 
taking
first the limit in which the lattice spacing  $a\to 0$, so that the
lattice densely fills the volume  $V$  ,and finally sending 
$V\to\infty$. 

In
QCD both limits exist . By renormalization group arguments the lattice
spacing  $a$, at a given value of the coupling constant  $g$, is 
related to
the physical momentum scale $\Lambda_L$ as
\begin{equation}
a \mathop\simeq_{g^2\to 0} \frac{1}{\Lambda_L} \exp(- \frac{b}{g^2})
  \label{eq22}\end{equation}
where  $b$ is minus the inverse of the first coefficient of the beta
function of the theory and is positive ,because of asymptotic 
freedom.
Eq.(22) will be valid at sufficiently small values of  $g^2$. 

At 
sufficiently
small  $g^2$  $a$  is small in physical units and the granularity of 
the lattice
becomes irrelevant.   

Moreover a mass gap exists in the theory (a non
zero minimum mass) ,which determines a finite correlation length  
$\xi$. If
the lattice is sufficiently large with respect to  $\xi$  the 
infinite volume
limit is reached. Lattice will be a good approximant to the continuum 
if
\[ a \ll \xi\ll L a\]
$L$ being the linear size of the lattice. Lattice calculations are in 
that
case an element of the sequence eq(12) near to the limit.Lattice will
also uniquely identify the ground state.

The symmetry of the ground state ,which is at the basis of the
confinement mechanism can safely be studied on the lattice.

\section{Feynman integral and perturbation theory.}
The only known computable functional  integral is the gaussian
integral,with a lagrangean quadratic in the fields,which corresponds 
to
free fields.

The action can be written as
\begin{equation}
S_0 = \frac{1}{2} \int dx dy \varphi_i(x) {\cal D}^{-1}_{ij}(x-y) 
\varphi_j(y)
 \label{eq23}\end{equation}
The resulting equations of motion are $\delta S/\delta\varphi(x) = 0$
or     $\int {\cal D}^{-1}_{ij}(x-y) \varphi_j(y) = 
0$                                         .
The matrix      $ {\cal D}^{-1}_{ij}(x-y)$  
is the inverse of the propagator. For the scalar field
\[  {\cal D}^{-1}_{ij}(x-y) = (\partial^2 + 
m^2)\delta_{ij}\delta(x-y)\] 
The discretized version for $S_0$ is
\[ S^0_L = \frac{1}{2} a^4 \sum_{ab} \varphi_a {\cal D}^{-1}_{ab} 
\varphi_b\]
 ${\cal D}^{-1}_{ab}$       is a symmetric matrix which can be 
diagonalized by an orthogonal
transformation
\begin{equation}
O_T {\cal D}^{-1} O = {\cal D}^{-1}_{diag}
 \label{eq24}\end{equation}
The jacobian of the transformation is equal to 1 so that
\begin{equation}
Z = \int \prod d\varphi_i e^{-\frac{1}{2}a^4\sum\varphi_i^2( {\cal 
D}^{-1}_{diag})_{ii}}
= c (\det {\cal D})^{1/2}
 \label{eq25}\end{equation}
The correlation functions are given by
\begin{equation}
\langle0|T(\varphi_{i_1}(x_1)\ldots\varphi_{i_n}(x_n)|0\rangle
= \frac{1}{Z}\int\prod d\varphi_i 
\varphi_{i_1}(x_1)\ldots\varphi_{i_n}(x_n) e^{-S_0}
 \label{eq26}\end{equation}
The integral is easy to compute,again by diagonalizing  ${\cal 
D}^{-1}_{ij}$   ,giving
\[ c (\det {\cal D})^{1/2} \sum_{perm} \prod {\cal D}_{i_a i_b}\]
The sum runs over all possible choices of the pairs $i_a,i_b$.

The only non zero correlators are the two point functions. The Hilbert
space is Fock space. The generic correlator is the product
of two point functions.

In the perturbative approach to quantization the action $S$ is split 
in a
quadratic part  $S_0$   and terms containing higher powers of the 
fields
   \[               S= S_0 + S_I\]
The functional integral is then computed by expanding the weight in
powers of $S_I$
\begin{equation}
Z = \int{\cal D}\varphi e^{-S_0 - S_I}
=  \int{\cal D}\varphi e^{-S_0}\sum_n \frac{(-1)^n}{n!} S_I^n
 \label{eq27}\end{equation}
$S_I$  is a polynomial in the fields.
The generic correlation function is then given by
\begin{equation}
\langle0|T(\varphi_{i_1}(x_1)\ldots\varphi_{i_n}(x_n)|0\rangle
= \frac{1}{Z}\int\prod d\varphi_i e^{-S_0}
\varphi_{i_1}(x_1)\ldots\varphi_{i_n}(x_n) 
\sum_k \frac{(-1)^k}{k!} S_I^k
 \label{eq28}\end{equation}
and again is a gaussian integral which can be computed.

The result, a 
sum
of products of propagators,  is nothing but the well known Wick's
theorem,and each term corresponds to a Feynman diagram.

Eq(26) is the
interchange of the order of two limits ,which is not always 
legitimate.

The perturbative expansion can be seen as the functional version of 
the
saddle point method for evaluating ordinary integrals.
  Consider a one variable integral of the form
\begin{equation}
I = \int_a^b dx\, \exp(i f(x)) g(x)
 \label{eq29}\end{equation}
If  $f$ and $g$  are analytic in a domain of the complex plane 
including the
real interval  $(a,b)$ the path of integration can be deformed in that
domain to include the points $\bar x$   where   the phase  $f(x)$ is
stationary,i.e.   $\frac{df}{dx} =0$.  The neighborhood of these 
points will give the
main contribution to the integral :far from it the phase factor is
rapidly oscillating ,and the contribution to the integral negligible.
The saddle point method consists in writing  $I$  in the form
\begin{equation}
I = \int_{a,C}^b d\delta\, e^{i f(\bar x) +\frac{i}{2} f''(\bar 
x)\delta^2} 
g(\bar x + \delta)  \sum_n \frac{i^n R^n(\delta)}{n!})
 \label{eq30}\end{equation} 
 where     $R(\delta)  = f - f(\bar x) - \frac{1}{2} f''(\bar 
x)\delta^2$                       
(the linear term drops at  $\bar x$ ), and
the factor multiplying the exponential is intended as a power series in
$\delta$. The integral is then gaussian . If more saddle points  
$\bar x_i$  exist $I$  will
be the sum of analogous expansions around $\bar x_i$ . 

The method 
works in
practice. No systematic control ,however,exists of the approximations
involved.

  For the Feynman integral the idea is the same, except that now 
there are
infinitely many integration variables. The phase is the action  $S$, 
the
integration variables are the fields, and a saddle point
\begin{equation}
\frac{\delta S}{\delta \varphi(x)} = 0
 \label{eq31}\end{equation} 
is nothing but a classical solution of the equations of motion.

The point  $\bar\varphi =0$ is certainly a saddle point. Around it 
one can expand the
action as
\begin{equation}
S[\varphi] = S[\bar\varphi] + \left. \frac{1}{2}
\frac{\delta^2 S}{\delta \varphi_i\delta\varphi_j}\right|_{\varphi = 
\bar\varphi}\delta_i\delta_j
+ S_I
 \label{eq32}\end{equation} 
For  $\bar\varphi=0$ $S[\bar\varphi]  =0$,            
$ \frac{1}{2}
\frac{\delta^2 S}{\delta \varphi_i\delta\varphi_j}\delta_i\delta_j = 
S_0$
and the saddle point formula is identical to
the perturbative expansion eq (27). 

In general, if different saddle 
points
exist,the expression for the partition function becomes
\[ Z = \sum_{\bar\varphi_i}
e^{-S[\bar\varphi_i]}
\int[d\delta] e^{-\frac{1}{2}\frac{\delta^2 S}{\delta 
\varphi_i\delta\varphi_j}\delta_i\delta_j}
\sum_n\frac{(-1)^n}{n!} S_I^n\]
and all the classical solutions with finite action contribute. 

Such solutions (if non trivial) are called instantons.
  In a non abelian gauge theory finiteness of the euclidean action 
means
\[\left| \int d^4x Tr\left[ G_{\mu\nu}(x) G_{\mu\nu}(x)\right]\right| 
< \infty\]
or that the field $G_{\mu\nu}$    decreases more rapidly than 
$1/r^2$     as   $r\to\infty$. The
fields $A_\mu$ at large distance are a pure gauge:  $A_\mu = i 
\partial_\mu U\, U^\dagger$         with $U$   an
element of the gauge group. A topological number exists which counts 
how
many times in this mapping the group is covered when the point at
infinity sweeps the sphere  $S_3$ . The topological charge is
\[ Q = \frac{1}{16\pi^2}  \int d^4x Tr\left[ G_{\mu\nu}(x) 
G^*_{\mu\nu}(x)\right]\]
  and is an integer. Explicit instanton solutions were found\cite{6}
which are self dual or antiselfdual
\[ G_{\mu\nu} = \pm G^*_{\mu\nu}\]
A non zero value of  $Q$  implies then a non zero value of  $G^2 =
G_{\mu\nu}G_{\mu\nu}$, or that
$\langle0| G^2|0\rangle\neq 0$,  a phenomenon which is known as gluon 
condensation.
 $\langle G^2(x)\rangle$,  which must be  $x$ independent by 
translation invariance, is known as
gluon condensate. A non zero value of  $\langle G^2\rangle$  is a 
highly non-perturbative
result .  Indeed renormalization group dictates for   $\langle 
G^2\rangle$ the form of an
operator of dimension 4,
\begin{equation}
\langle G^2\rangle \simeq e^{-4/b_0 g^2}
 \label{eq33}\end{equation} 
where   $b_0 > 0$ is proportional to the first non zero coefficient 
of the beta
function.  An expression like eq(33) is non analytic at  $g^2=0$, and 
therefore
non expandable in a power series of  $g^2$ . A complete 
classification of the
instantonic solutions does not exist.Models have been developed based 
on
quasi-solutions,consisting of ensembles of single instantons.If their
average distance is large compared to their size the ensemble is 
called
an instanton gas\cite{7}, if the overlapping is important the ensemble is 
called an
instanton liquid\cite{8}. Attempts to improve perturbative expansion by 
adding
to the perturbative saddle point the saddle points corresponding to
instantons have been done during the years,starting from the pioneering
approach of ref[7].  Although useful to describe chiral properties,
instantons fail in accounting for confinement of color. 

Instantons 
were at
the basis of the SVZ\cite{9} sum rules of QCD: that approach, which will be
described in next section, is based on the existence of 
condensates, the
original idea being that instantons provide a slowly varying 
background
field for quantum fluctuations.
\section{Condensates.}
The time ordered product of two currents $j_\mu(x),j_\nu(0)$,     
e.g. electromagnetic
currents,can be written at short distances as a sum of local operators
times c-number coefficients\cite{10}.
\begin{equation}
T(j_\mu(x) j_\nu(0)) \simeq
C^0_{\mu\nu}(x) I + C^4_{\mu\nu}(x) G^a_{\mu\nu} G^a_{\mu\nu}
+ C^\psi_{\mu\nu}(x) m \bar\psi\psi + \ldots
 \label{eq34}\end{equation} 
The operators in eq(34) are ordered by increasing order of 
dimension in mass.
The expansion is rigorously valid in perturbation theory, but is assumed to 
hold
also when perturbation theory is not expected to work. Take the vev of eq.(34) 
and
Fourier transform. The left side gives
\begin{equation}
\Pi_{\mu\nu}(q) \equiv\int e^{iqx}
\langle 0|T(j_\mu(x) j_\nu(0))|0\rangle
d^4x = (q_\mu q_\nu - q^2 g_{\mu\nu})\Pi(q^2)
  \label{eq34}\end{equation} 
the tensor structure being dictated by gauge invariance.

The spectral
representation reads
\begin{equation}
\Pi(q^2) = \Pi(0) -
q^2\int_{4 m_\pi^2}^\infty \frac{ds}{s}\frac{R(s)}{s - q^2 + 
i\varepsilon}
  \label{eq36}\end{equation} 
with         
\[ R(s) = \frac{\sigma_{e^+e^-\to hadrons}}{ 
\sigma_{e^+e^-\to\mu^+\mu^-}}\]                     
The right side of eq.(34) gives
\begin{equation}
(q_\mu q_\nu - q^2 g_{\mu\nu})\left\{
C_0(q^2) + C_g(q^2) G_2 + C_\psi(q^2) \langle m \bar\psi\psi\rangle + 
\ldots\right\}
  \label{eq37}\end{equation} 
The term $C_0(q)$  corresponds to the usual perturbative expansion 
and is
constant modulo log's. The other terms, known as "higher twists",
are non
perturbative in nature.

  The approximate equality
\begin{equation}
-q^2\int_{4 m_\pi^2}^\infty \frac{ds}{s}\frac{R(s)}{s - q^2 + 
i\varepsilon}
\simeq
\left\{
C_0(q^2) + C_g(q^2) G_2 + C_\psi(q^2) \langle m \bar\psi\psi\rangle + 
\ldots\right\}
  \label{eq38}\end{equation} 
can be exploited by appropriate weighting which emphasises the region
where the equality is a good approximation on the average:the 
procedure
is known as sum rules\cite{9}.

The dispersion integral in eq (38) is evaluated 
by
inserting for $R(s)$ the resonances at low s and a constant above some
threshold. A nice phenomenology results,relating resonance parameters 
to
the condensates,which can be determined consistently giving\cite{9}
\[ \begin{array}{ll}
G_2 &= 0.012\,{\rm GeV}^4\\
\langle\bar\psi\psi\rangle &= - 0.13\,{\rm GeV}^3
\end{array}\]
A more recent determination for $G_2$  is\cite{11}
\[ G_2 = (0.024\pm 0.011)\,{\rm GeV}^4\]
\section{Lattice determination of the condensates.}
Consider the gauge invariant correlator
\begin{equation}
{\cal D}_{\mu\nu,\rho\sigma}^C(x) =
\langle0| Tr\left\{
G_{\mu\nu}(x) S_C(x,0) G_{\rho\sigma}(0) 
S_C^\dagger(x,0)\right\}|0\rangle
 \label{eq39}\end{equation}
where  $G_{\mu\nu} = \sum_a\frac{\lambda^a}{2} G^a_{\mu\nu}$, $A_\mu 
= \sum_a\frac{\lambda^a}{2}A^a_\mu$
  and  $\lambda^a$  are the generators of the gauge group in the
fundamental representation.
\[ S_C(x,0) = P\left\{ \exp(i \int_{0,C}^x A_\mu(x) dx^\mu)\right\}\]
is the parallel transport from  0  to $x$  along the path  $C$. Under 
a gauge
transformation  $U(y)$, 
$ S_C(x,0) \to U(x) S_C U^\dagger(0)$, which makes ${\cal 
D}_{\mu\nu,\rho\sigma}^C$ 
 gauge invariant. In what
follows we will chose for $C$ the straight line connecting 0 and $x$.
${\cal D}_{\mu\nu,\rho\sigma}^C$ can be considered as the split point 
regulator a la Schwinger
of the gluon condensate  $G_2$. 
The Wilson operator product expansion gives indeed

\begin{equation}
{\cal D}_{\mu\nu,\rho\sigma}^C(x) \mathop\simeq_{x\to0}
\frac{c_1}{x^4} \langle I\rangle + c_4 G_2
\label{eq40}\end{equation}
whence $G_2$   can be extracted. Poincare' invariance implies\cite{12}
\begin{eqnarray}
{\cal D}_{\mu\nu,\rho\sigma}(x)&=& (g_{\mu\rho} g_{\nu\sigma} -
g_{\mu\sigma} g_{\nu\rho})[ D(x) + D_1(x)] + \label{eq41}\\
&&+ (x_\mu x_\rho g_{\nu\sigma}
- x_\nu x_\rho g_{\mu\sigma} - x_\mu x_\sigma g_{\nu\rho} +
x_\nu x_{\sigma} g_{\mu\rho}) \frac{\partial D_1}{\partial x^2}\nonumber
\end{eqnarray}
or,choosing the $x^0$  axis along $x^\mu$   one can define\cite{13}
\begin{eqnarray*}
{\cal D}_{||} &=& \frac{1}{3}\sum_{i=1}^3 {\cal D}_{0i,0i}
= D + D_1 + x^2 \frac{\partial D_1}{\partial x^2}\\
{\cal D}_{\perp} &=& \frac{1}{3}\sum_{i,j} {\cal D}_{ij,ij} =
D + D_1
\end{eqnarray*}
On the lattice,which is a formulation in terms of parallel 
transports, the
correlators are easy to implement: the field strength is the open
plaquette,
the parallel transport the product of elementary links.
The correlator is measured as the average on the ensemble 
of
the configurations produced by MonteCarlo simulations of the operator:
\vskip\baselineskip
\par\noindent
\begin{minipage}{0.6\linewidth}
${\cal D}^L_{\mu\nu\rho\sigma}=\left\langle
\hbox{
\includegraphics[width = 0.4\linewidth]{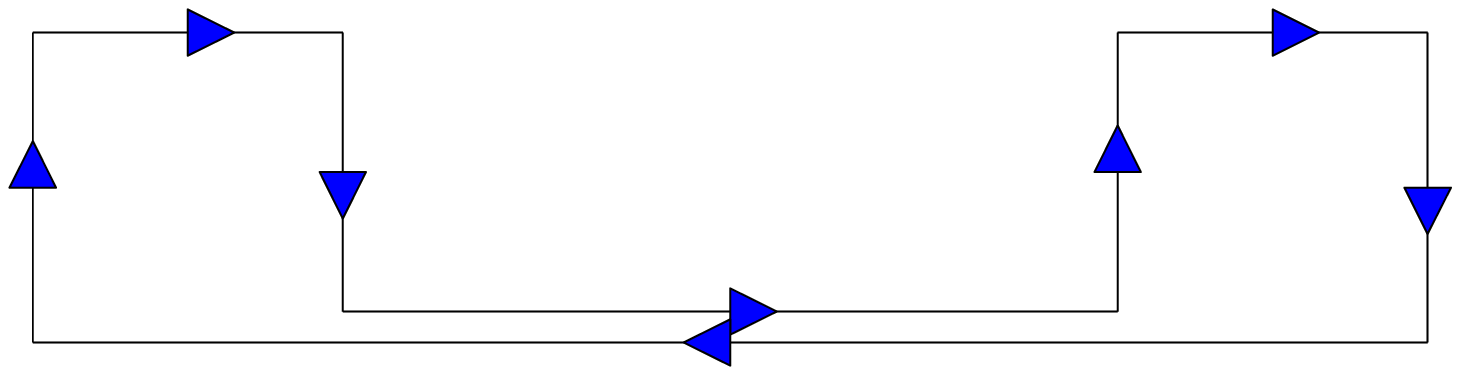}
}
\right.
$
\end{minipage}
\begin{minipage}{0.39\linewidth}\hskip-30pt
$-\frac{1}{N_c}\left.
\hbox{
\includegraphics[width = 0.5\linewidth]{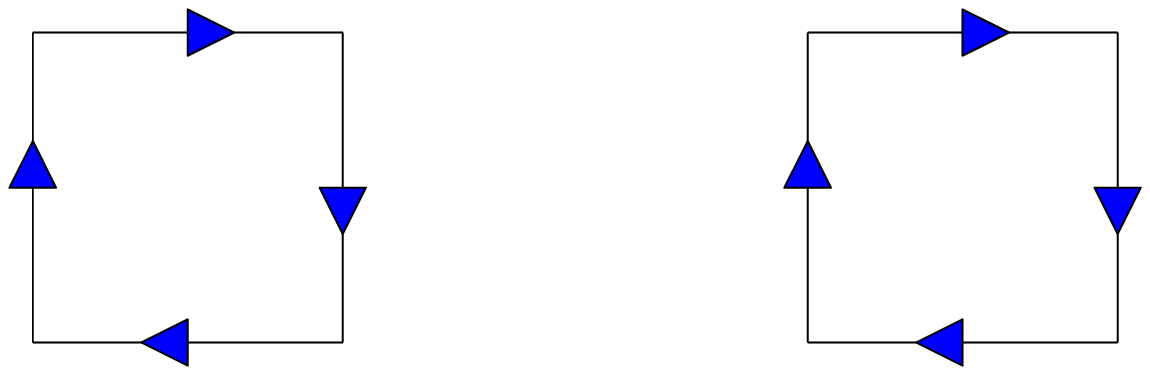}}\right\rangle$
\end{minipage}\par\noindent
\begin{minipage}{\linewidth}
\vskip-10pt\hskip60pt\hbox{${P}_{\mu\nu}$}
\hskip54pt${P}_{\rho\sigma}$
\hskip95pt${P}_{\mu\nu}$
\hskip40pt${P}_{\rho\sigma}$
\end{minipage}
\vskip\baselineskip

The figure represents the product of elementary links.
  The correlators $D$     and    
$ D_1$ are plotted versus the distance in
fm in fig.~1. 

\begin{figure}[htb]
\includegraphics[width = 0.9\linewidth]{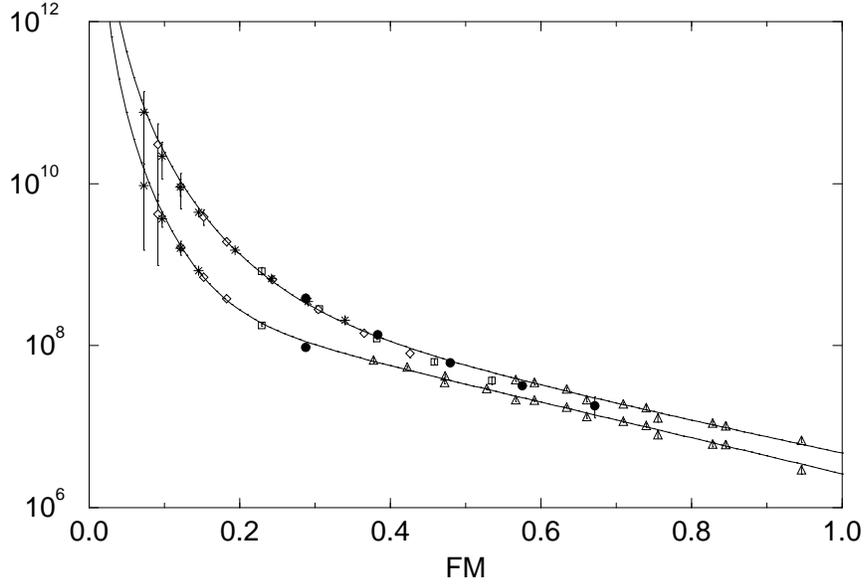}
\caption{$D$ (upper curve) and $D_1$ vs distance in fm, as determined on lattice\cite{13,14}.}
\end{figure}

The lines represent a best fit of the form
\begin{eqnarray}
D &=& A e^{-x/\lambda} + \frac{a}{|x|^4}\label{eq42}\\
D_1 &=& B e^{-x/\lambda} + \frac{b}{|x|^4}
\nonumber
\end{eqnarray}
which corresponds to the OPE  eq.(40).
One can extract $A$, $B$  and the correlation length  $\lambda$. 
$G_2$  is proportional,
with known coefficients  to  $A+B$.
The result is
\[ \begin{array}{lll}
\hbox{ Quenched\cite{15}}\hskip0.1in SU(2) & G_2 = (.33\pm.01)\,{\rm GeV}^4 & \lambda = 
.16\pm .02\,{\rm fm}\\
\hbox{ Quenched\cite{13,14}}\hskip0.1 in SU(3) & G_2 = (.15\pm.03)\,{\rm GeV}^4
& \lambda = .22\pm.02\,{\rm fm}\\
\hbox{Full QCD, extrapolated}  & G_2 = 
(.022\pm.005)\,{\rm GeV}^4
& \lambda = .34\pm.03\,{\rm fm}\\
\hbox{to physical q masses\cite{16}} & &
\end{array}\]
The value of $G_2$ for full QCD agrees with recent phenomenological determinations\cite{11}.
The correlation length is definitely too small in quenched theories, 
both
SU(2) and SU(3) to consider the background field as slowly varying,
it is slightly larger in full QCD.
 
By similar techniques the chiral condensate can be extracted from 
the
correlator\cite{17}
\[ {\cal D}_\psi(x) = m \langle\bar\psi(x) S_C(x,0) \psi(0)\rangle\]
which allows the OPE
\[ {\cal D}_\psi(x)\simeq \frac{C_1}{|x|^2} + C_2 \langle m 
\bar\psi\psi\rangle\]
\section{The stochastic vacuum.}
In the Schwinger gauge   $(x-z_0)^\mu A_\mu(x) = 0$, the gauge field 
$A_\mu$  can be expressed in
terms of $G_{\mu\nu}$
\begin{equation}
A_\mu^a(x) =
\int \alpha d\alpha (x-z_0) G^a_{\mu\nu}(z_0 + \alpha(x-z_0))
\label{eq43}\end{equation} 
and any physical quantity can be expressed in terms of correlators 
of
field strengths in the vacuum. A cluster expansion is made, and the
assumption that higher correlators are products of two point
correlators, higher clusters being negligible (stochastic vacuum model). The two point 
correlators
determined on the lattice are used as an input, and several physical
quantities can be computed: the heavy $q\bar q$  bound states or high 
energy
cross sections. A successfull phenomenology results\cite{18}.

We have shown that non perturbative effects, like gluon condensate, are
important in QCD and can be studied on the lattice. In the next 
lecture we
shall investigate what symmetry of the vacuum is at the origin of 
color
confinement.
\section{Confinement of color.}

A deconfinement transition is observed in lattice simulations of QCD.

The static thermodynamics of a field system at temperature  T  is
described by the partition function  
\[ Z = tr\left\{ e^{- H/T}\right\}\]
It is a
known theorem that $Z$ is given by a Feynman euclidean integral, 
extending
on the  3-dimensional physical space and on the time interval 
$(0,1/T)$,
with periodic boundary conditions for bosons,antiperiodic for 
fermions.

On the lattice  $Z$  is obtained by simulating the theory on a lattice
$N_S^3\times N_T$ ,with $N_S \gg N_T$. In terms of the lattice 
spacing $a$  the
temperature is then given by  $T =1/N_T a$. The only free parameter 
in the
simulation is the gauge coupling constant $g^2$    or better the 
parameter
 $\beta = 2 N_c/g^2$. In terms of $\beta$ the lattice spacing can be 
written as [see
eq(22)] 
\[ a = \frac{1}{\Lambda} \exp(-\frac{\beta}{b})\]
or
\begin{equation}
T = \frac{1}{N_T a} = 
\frac{\Lambda}{N_T} \exp(\frac{\beta}{b})
\label{eq44}\end{equation}
Low temperature corresponds to small $\beta$ (strong coupling), high 
temperature
to large $\beta$ (weak coupling). 

This is the opposite to what 
happens in
ordinary spin systems, where the temperature plays the role of 
coupling
constant, and is due to asymptotic freedom.
 Confinement, for pure gauge theories, is related to the Polyakov
line, which is the parallel transport along the time (temperature) 
axis
from 0 to N, closed by periodic boundary conditions. The vev of the Polyakov line $\langle 
L\rangle$ can be
interpreted as $\exp(-\mu_q)$, $\mu_q$ being the chemical potential 
of an isolated
quark. In the confined phase $\mu_q=\infty$, $\langle L\rangle = 0$, while one can 
have
$\langle L\rangle\neq 0$
 in the
deconfined phase. $\langle L\rangle$   can be called an order 
parameter for confinement,
the symmetry being $Z_{N}$.
 What is usually determined on the lattice\cite{19} is the correlator of two
Polyakov lines, which by cluster property behaves at large distances as
\begin{equation}
G(d) = \langle L(\vec x) \bar L(\vec y)\rangle \simeq
A \exp(-\sigma d a N_T) + |\langle L\rangle|^2
\label{eq45}\end{equation}
with $d = |\vec x-\vec y|$.

The energy of two static quarks at distance  $d$  is related to 
$G(d)$ by the
equation
\begin{equation}
V(d) = - \frac{1}{a N_T} \log( G(d))
\label{eq46}\end{equation}
What is found on the lattice is that a temperature $T_c$ exists such 
that
\[\begin{array}{llll}
T < T_c \hskip0.15in& |\langle L\rangle| = 0 \hskip0.15in & V(d) 
\mathop\simeq_{d\to\infty}  
\sigma d
\hskip0.15in &\hbox{(Confinement)}\\
T > T_c & |\langle L\rangle| \neq 0 & V(d) \mathop\simeq_{d\to\infty} 
{\rm const.} &\hbox{(Deconfinement)}
\end{array}\]
For quenched $SU(2)$ one finds $T_c/\sqrt{\sigma} \simeq 0.7$.

A finite size scaling analysis of $G$  at different spatial  sizes  
$N_S$,
allows to extract the  critical index $\nu$ The result is  $\nu \simeq 0.62$.
The phase
transition belongs to the same class of universality as the  3d ising
model,as expected\cite{20}. For quenched $SU(3)$ the transition is found to 
be
weak first order, $\nu  =1/3$ and  $T_c/\sqrt{\sigma} =0.65$  which 
means  
$T_c  =270$~Mev if the
conventional value  $\sqrt{\sigma} = 425$~MeV is assumed for the 
string tension.

In the presence of quarks $\langle L\rangle$ cannot be an order 
parameter, since the
symmetry $Z_{N_c}$ is explicitely broken by the very presence of the
quarks. At  $m_q   =0$, the chiral limit, an order parameter is the 
chiral
condensate $\langle\bar\psi\psi\rangle$: however 
in reality
the chiral symmetry 
is explicitely broken  by quark masses. 
This variety of order parameters contrasts with the ideas
of the limit  $N_c\to\infty$. 

The idea is that the number of colors 
$N_c$ can be
considered as a parameter of the theory. In the limit $N_c\to\infty$  
with
$g^2 N_c =\lambda$ fixed, a theory is defined , which differs little 
from the theory at
finite $N_c$, say $N_c=3$.

In this spirit the mechanism of confinement should be fixed by the
limiting theory ,and be essentially the same at all values of  $N_c$ .

Quark loops being non leading in the expansion ,the mechanism of
confinement should be the same in full QCD and in quenched theory.

The
symmetry of the vacuum which is behind confinement, as explained
in sect (1) should be $N_c$   independent. In this respect the existing
situation for the order parameters looks rather confusing. 

\begin{figure}[htb]
\includegraphics[width = 0.9\linewidth,angle = 270]{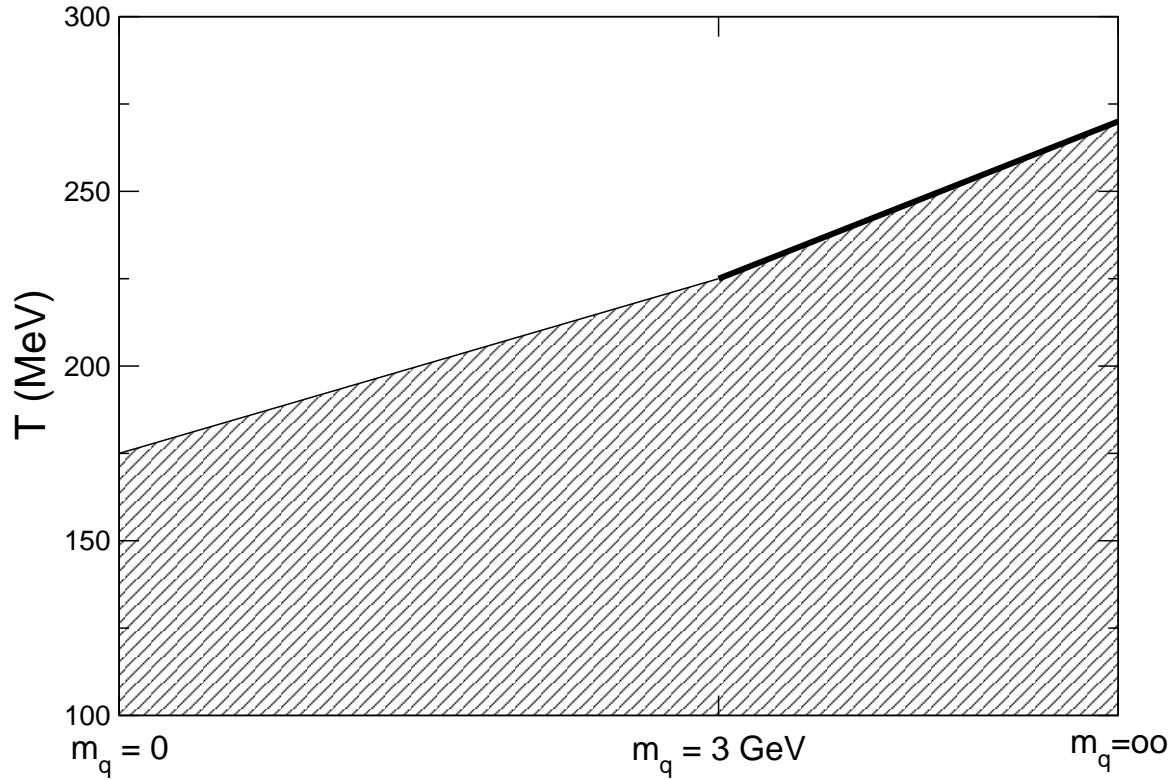}
\caption{A schematic phase dyagram for 2 flavour QCD. Shaded area is
confined, upper area deconfined.}
\end{figure}

Fig.2 shows
schematically the phase dyagram for 2-flavor QCD as a function of the
quark masses. The border between confined and deconfined phases is 
defined
as the simultaneous maximum of various susceptibilities\cite{21}.
>From $m_q = \infty$   to  $m_q = 3$~GeV  the transition is first 
order, and the
Polyakov line works as an order parameter as in quenched
QCD. At  $m_q  =0$ there are arguments that the
transition is second order\cite{22}. In the central region none of
the susceptibilities which have been determined  ($\bar\psi\psi$,
susceptibility, derivative of the specific heat) goes
large 
at the border between the two regions
as the volume goes to infinity, and the conclusion is
that maybe there is no transition, but only a crossover.

Still, if the deconfining transition is a change of
symmetry,as argued in sect.1, there must exist an order(or
disorder) parameter which distinguishes confined from
deconfined, possibly the same as in quenched theory,in the
line of $N_c\to\infty$ ideas.

\section{Duality}
The symmetry responsible for confinement has to be a
symmetry of the (strong coupling) confining phase, which, in
the langauge of statistical mechanics, is disordered. The natural 
question
is: what can be the symmetry of a disordered phase? 
The answer is in a
known concept in statistical mechanics, namely
duality\cite{23}. $d$ dimensional systems admitting non local configurations 
with non
trivial topology  in $d-1$ dimensions, can have two complementary
descriptions. 

One description in terms of the ordinary local
fields $\phi$,in which topological excitations $\mu$    are non local,
and $\langle\phi\rangle$   are the order parameters; this description 
is
convenient in the weak coupling regime  $g \ll 1$     (ordered phase).

The other one ,dual description,in which the excitations $\mu$
become local fields ,their  vev's $\langle\mu\rangle$  are the order
parameters: the coupling constant $g_D$ in this description is
related to the original $g$  as $g_D\sim 1/g$. Duality maps the
strong coupling regime of the direct description into the weak
coupling regime of the dual. 

A number of systems admitting dual
descriptions are known in the literature.The prototype system
is the 2d Ising model\cite{24}, in which the dual excitations are
kinks, and the dual description is again a  2d ising model
with $g_D\sim 1/g$; the N=2 SUSY QCD  of ref\cite{25}, in which dual
excitations are monopoles; the  3d XY model\cite{26} where the dual
excitations are vortices,and the  Heisenberg magnet in 3d,
where the dual excitations are 2d Weiss domains\cite{27}.
In QCD the dual excitations are not exactly known, but important
information on their properties exists as we shall see below. 

Two main candidates
were originally proposed: 
\begin{itemize}
\item[a)] monopoles\cite{28,29}. Their condensation
in the confining vacuum generates magnetic superconductivity,in
the same way as the condensation of Cooper pairs generates
ordinary superconductivity.

The chromoelectric field between a
$q \bar q$  pair is channeled into electric Abrikosov flux tubes, so
that energy is proportional to the distance.
\item[b)] $Z_N$  vortices\cite{30}
\end{itemize}
In what follows we shall analyze the option a). As for vortices we
refer to\cite{31,32} and to references therein.

\section{Basic superconductivity\cite{33}.}
Before looking for dual (magnetic) superconductivity,we recall the
basic concepts of ordinary superconductivity. The Landau-Ginzburg\cite{34}
density of free energy (effective lagrangean) is determined by
arguments of symmetry and of scale.
\begin{equation}
{\cal L} = -\frac{1}{4} F_{\mu\nu} F_{\mu\nu} +
\frac{1}{2} (D_\mu\phi)^* D_\mu\phi + \frac{m^2}{2}\phi^*\phi 
-\frac{\lambda}{4}
(\phi^*\phi)^2 + \hbox{irrelevant terms}
\label{eq47}\end{equation}
$\phi$
 is a charged scalar field describing Cooper pairs ,  $D_\mu = 
\partial_\mu - i q A_\mu$      is the
covariant derivative and the irrelevant terms are those of dimension
higher than 4.  $m^2 = m^2(T)$ and $\lambda = \lambda(T)$ depend on 
the temperature.

For $ T< T_c$
 $m^2   > 0$  and the potential has a mexican hat shape as a function 
of $R(\phi)$
an of $Im(\phi)$. 

For $T> T_c$ $m^2 < 0$ and the minimum of the 
potential is at  
$\phi = 0$.
If we parametrize   $\phi = \rho e^{i q \theta}$, then
 $\tilde A_\mu = A_\mu - \partial_\mu\theta$ is a gauge invariant 
quantity: indeed under a gauge transformation $A_\mu \to A_\mu + 
\partial_\mu\Lambda$, $\theta\to\theta + \Lambda$.
In terms of the new variables
\begin{equation}
{\cal L} = -\frac{1}{4} F_{\mu\nu} F_{\mu\nu} +\frac{q^2\rho^2}{2}
\tilde A_\mu\tilde A_\mu +\frac{1}{2}(\partial_\mu\rho)^2 +
\frac{m^2}{2}\rho^2 -\frac{\lambda}{4}\rho^4
\label{eq48}\end{equation}
A static homogeneous solution is $\rho =\bar\rho$, a constant. The effective lagrangean for the
photon is in that case
\[ {\cal L} = -\frac{1}{4} F_{\mu\nu}F_{\mu\nu} +  q^2\bar\rho^2
\tilde A_\mu\tilde A_\mu \]
The equation of motion is
\begin{equation}
\vec\nabla\wedge\vec H + q^2\rho^2 \vec{\tilde A} = 0
\label{eq49}\end{equation}
Taking the curl of both sides of eq(49)
 \begin{equation}
\nabla^2\vec H - q^2\rho^2 \vec H = 0
\label{eq50}\end{equation}
The second term in eq (49) is a gauge invariant current,and is known
as London current.

A non zero current with zero electric field means
zero resistivity or superconductivity.

Eq(50) means that the magnetic
field has a finite penetration depth in the superconductor , $1/q\rho$,  
and
this is nothing but Meissner effect. If $m \gg q\bar\rho$ the 
superconductor
is named Type II. In that case, when trying to introduce a magnetic
field in the bulk, a penetration through separated Abrikosov flux
tubes is energetically favoured. 

Outside the tube $\tilde A=0$   and 
$\oint_C\vec{\tilde A}d\vec x = 0$, $C$
being a path which encercles the flux tube. Since $\vec{\tilde A} = 
\vec A -
\vec\nabla\phi$, this means
 \begin{equation}
\int_C \vec A\,d\vec x = \int_C \vec\nabla\,d\vec x = \frac{2\pi n}{q}
\label{eq51}\end{equation}
The magnetic flux in the tube is quantized (Dirac quantization
condition). 
A flux tube is to all effects a monopole antimonopole pair
sitting at the ends with energy proportional to the length.

The order parameter of the system is $\Phi = q\bar\rho$, the vev of a
charged field.

\section{Monopoles in QCD.}
Monopoles are always abelian\cite{35}: the magnetic monopole term in the
multipole expansion of the field produced at large distances by any
hadronic matter distribution ,obeys abelian field equations,and can
always be reduced by a gauge transformation to the form $(0,0, a/r)$
in polar coordinates, with
 \begin{equation}
a_\varphi = Q(1-\cos\theta)
\label{eq52}\end{equation}
and the Dirac string along the north pole. The  $N_c \times N_c$ matrix 
$Q$ obeys
the condition $e^{ig 4\pi Q} = 1$,
implying that,in the representation in which $Q$ is
diagonal, it has integer or half integer eigenvalues in units $1/g$.
A monopole is identified by a constant diagonal matrix of the algebra 
with
integer or half integer matrix elements: there exist $N_c - 1$ 
independent
monopole charges.  

The same result follows  from the procedure known 
as
abelian projection\cite{28}. We shall illustrate it  for the case of SU(2)
gauge group: generalization to arbitrary $N_c$ is straightforward.
Let  be any field in the adjoint representation. Define $\hat\phi 
=\vec\phi(x)/
|\vec\phi(x)|$, a
direction in color space, and
 \begin{equation}
F_{\mu\nu}= \hat\phi\cdot\vec G_{\mu\nu} -\frac{1}{g} \hat\phi\left(
D_\mu\hat\phi\wedge D_\nu\hat\phi\right)
\label{eq53}\end{equation}
Both terms in the expression (53) are color singlets and gauge
invariant.The combination is chosen in such a way that  bilinear terms
$A_\mu a_\nu$ and $A_\mu \partial_\nu\phi$  cancel. Indeed, by 
explicit computation
 \begin{equation}
F_{\mu\nu}= \hat\phi(\partial_\mu \vec A_\nu - \partial_\nu\vec 
A_\mu) -
\frac{1}{g}\hat\phi(\partial_\mu\hat\phi\wedge\partial_\nu\hat\phi)
\label{eq54}\end{equation}
A gauge transformation which brings $\hat\phi$ to a constant, e.g. to 
$(0,0,1)$
is called an abelian projection on $\vec\phi(x)$. In that gauge indeed
 \begin{equation}
F_{\mu\nu}= \partial_\mu(\hat\phi\vec A_\nu) - 
\partial_\nu(\hat\phi\vec A_\mu)
\label{eq55}\end{equation}
is an abelian field. If, in the usual notation   $F^*_{\mu\nu} =
\varepsilon_{\mu\nu\rho\sigma} F_{\rho\sigma}$, a
magnetic current can be defined as
\begin{equation}
j_\nu = \partial_\mu F^*_{\mu\nu}
\label{eq56}\end{equation}
This current is  identically conserved, because of the antisymmetry of $F^*_{\mu\nu}$
 \begin{equation}
\partial_\nu j_\nu = 0
\label{eq57}\end{equation}
and identifies a $U(1)$ magnetic symmetry of the system. 

In the
usual continuum formulation $j_\nu$ is zero (Bianchi identities). 

In a
compact formulation, like lattice, $j_\nu$   can be non zero. 

The 
U(1) magnetic
symmetry can be either realized \`a la Wigner, and then the Hilbert 
space
consists of superselected sectors with definite magnetic charge, or 
can be
Higgs broken. In the first case the vev $\langle\mu\rangle =
\langle0|\mu|0\rangle$  of any operator $\mu$   carrying
magnetic charge is strictly zero. In the second case there exists some
 $\mu$    such that  $\langle\mu\rangle\neq0$. 

The effective action 
for  
$\langle\mu\rangle$
 will then be of the
form  eq(47) and the system 
behaves as
a dual superconductor. The operator 
$\mu$  is a
disorder operator
\begin{center}
\begin{tabular}{ll}
  $\langle\mu\rangle\neq 0$ & in the disordered,confining phase \\
   $\langle\mu\rangle = 0$ & in the ordered, deconfined phase
\end{tabular}
\end{center}
A magnetically charged operator $\mu$ can be constructed,which is
magnetically charged in any given abelian projection\cite{36,37}.
The basic principle to construct  $\mu$ is the well known formula for
translations
\begin{equation}
e^{ipa}|x\rangle = |x+a\rangle\label{eq58}\end{equation}
In field theory the field $\phi(y)$ is the analog of  $x$. Its 
conjugate momentum
$\Pi(y)$ the analog of $p$, and the operator
\begin{equation}
\mu(\vec x,t) =
\exp\left[i
\int d^3\vec y\,\Pi(\vec y,t) \tilde\phi(\vec x-\vec y)\right]
\label{eq59}\end{equation}
gives
\begin{equation}
\mu(\vec x,t) |\phi(\vec y,t)\rangle = 
|\phi(\vec y,t) + \tilde\phi(\vec x-\vec y)\rangle
\label{eq60}\end{equation}
i.e. it adds the classical configuration  $\tilde\phi(\vec x-\vec y)$   to any field 
configuration.

For monopoles the field is  $\vec A(\vec x,t)$, the conjugate 
momentum $\partial_0\vec A(\vec x,t) = \vec E(\vec x,t)$, and the
classical configuration to be added is the vector potential generated 
at
the point $\vec y$   by a monopole sitting at the point  $\vec x$, e.g.
\[ \vec b(\vec x-\vec y) = -\frac{m}{2e}
\frac{\vec n\wedge\vec r}{r(r-\vec n\vec r)}\]
where the Dirac string has been put along the $\vec n$  axis.
The definition (60) has to be adapted to compact formulation .The 
details
are published in ref\cite{37}.
The result is of the form
\begin{equation}
\langle\mu(\vec x,t) \rangle =
\frac{\tilde Z(\beta,\vec x,t)}{Z(\beta)}
\qquad \beta = \frac{2 N_c}{g^2}
\label{eq61}\end{equation}
where $Z(\beta )$ is the partition function  
\[ Z(\beta) = \int[{\cal D}A][{\cal D}\psi][{\cal D}\bar\psi]
\exp(-\beta S_E)\]
and
\[ \tilde Z(\beta) = \int [{\cal D}A][{\cal D}\psi][{\cal D}\bar\psi]
\exp(-\beta \tilde S_E)
\]
$\tilde S_E$    differs from $S_E$  by
the introduction of a dislocation on the slice
$x^0=t$.  $\Delta S$  is proportional to the spatial volume,
and therefore fluctuates as  $N_S^{3/2}$, so that $\tilde Z$
fluctuates as $\exp(N_S^{3/2})$. It proves convenient to define
instead
\begin{equation}
\rho(\beta) \equiv \frac{d}{d\beta}\log\langle\mu\rangle =
\langle S\rangle_S - \langle S+\Delta S\rangle_{ S+\Delta S}
\label{eq62}\end{equation}
which is easier to measure.In terms of $\rho$
\begin{equation}
\langle\mu\rangle =
\exp\int_0^\beta\rho(\tau) d\tau
\label{eq63}\end{equation}
Fig.3  shows schematically the shape of $\langle\mu\rangle$   as a 
function of $\beta$, and that of $\rho$ .
\vspace{0.5truecm}
\begin{figure}[htb]
\includegraphics[width = 0.45\linewidth]{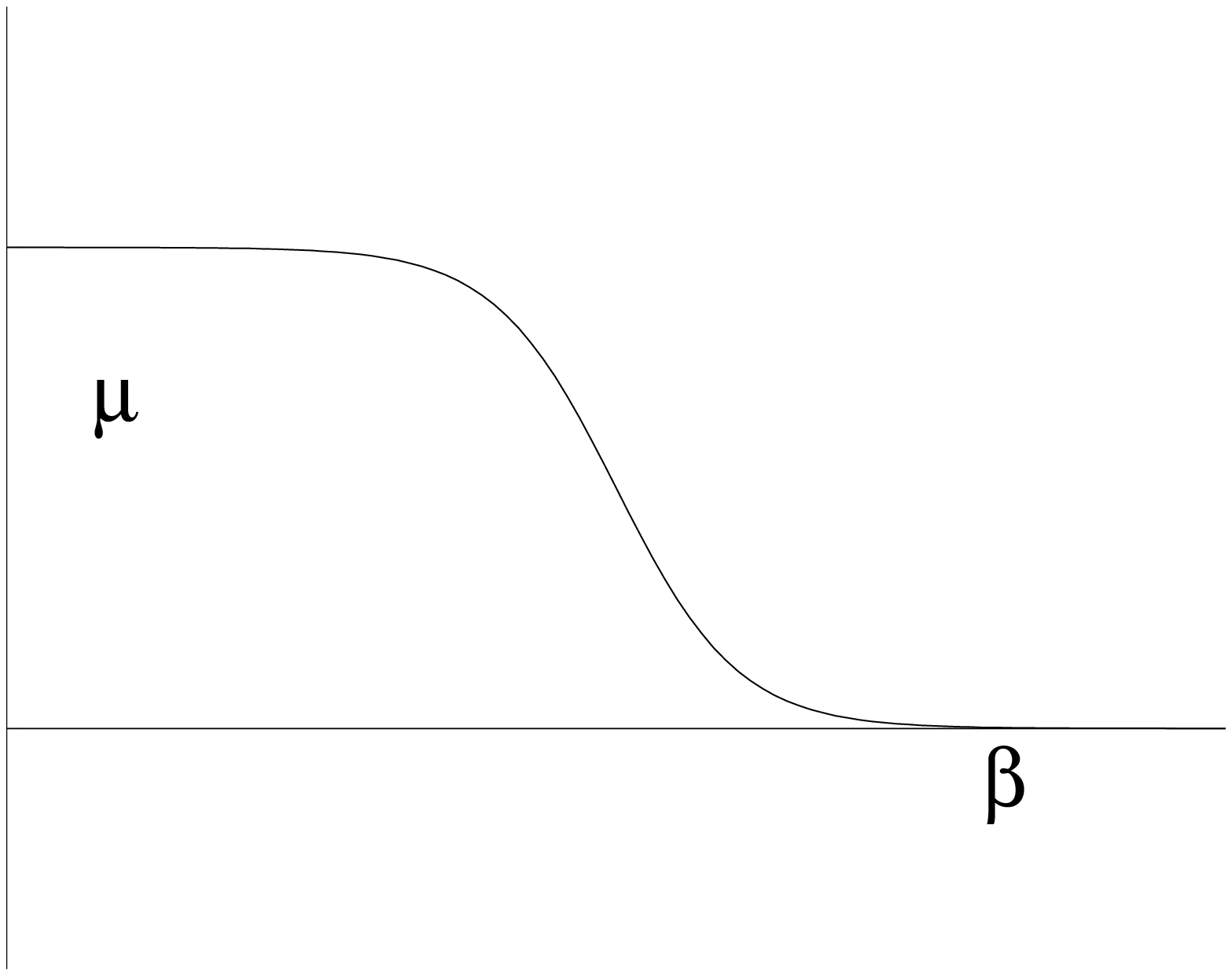}
\includegraphics[width = 0.45\linewidth]{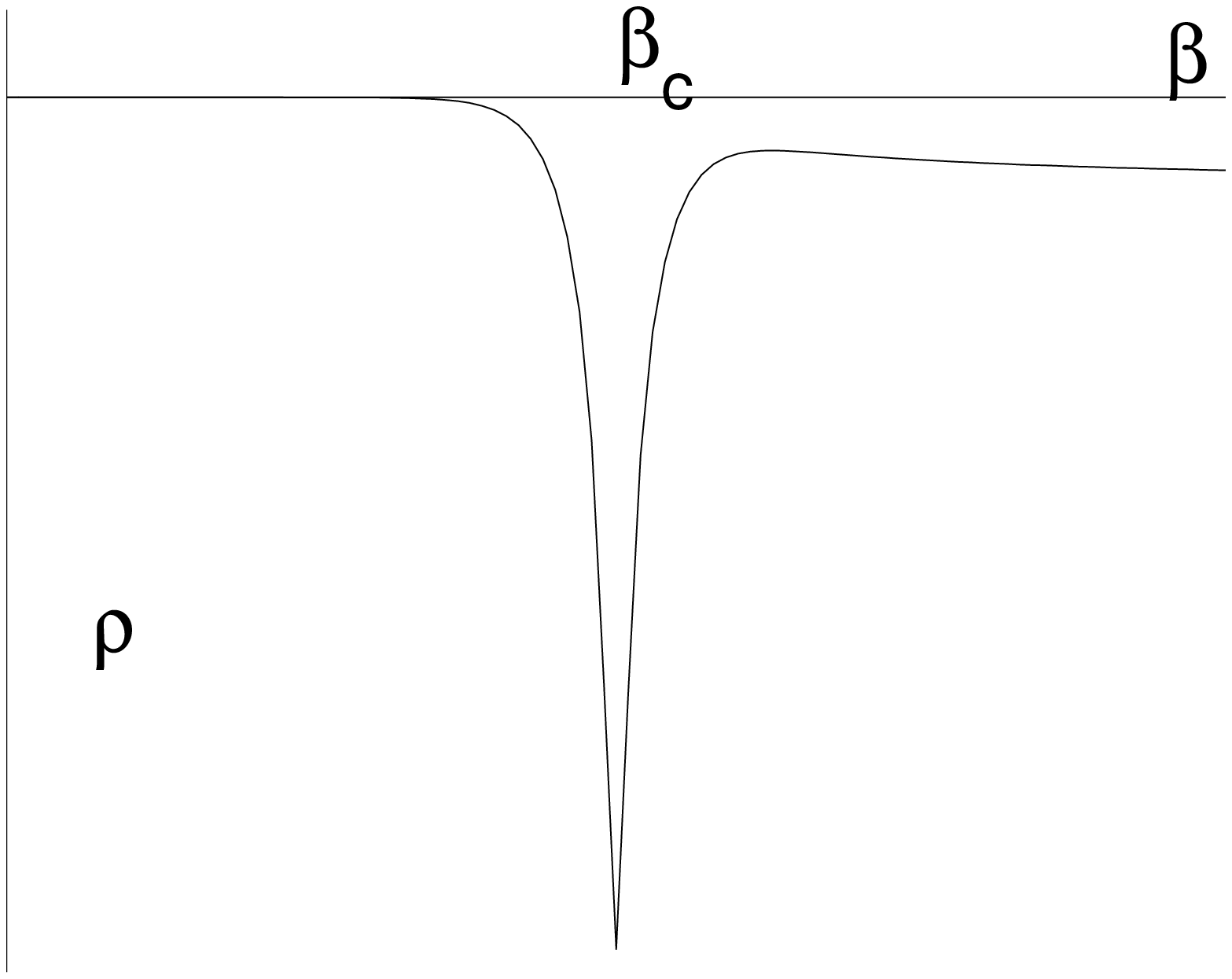}
\caption{Typical dependence on $\beta$ of  $\mu$ and $\rho$.}
\end{figure}
\vspace{0.5truecm}

For  $\beta <\beta_c$    (confinement) $\langle\mu\rangle\neq 0$, 
for  
$\beta > \beta_c\qquad  \langle\mu\rangle = 0$, at 
$\beta\sim\beta_c$                         
\[ \langle\mu\rangle 
\sim(\beta-\beta_c)^\delta \equiv (1-\frac{T}{T_c})^\delta\]
For finite lattices $\langle\mu\rangle$  cannot be strictly zero 
above 
$\beta_c$  ,because it is
analytic in $\beta$  and therefore if it were zero on a line it would 
be
 identically zero. Only in the infinite volume limit singularities 
can develop\cite{39}
 and   can be identically zero above $\beta_c$.  The limit 
$N_s\to\infty$
 can be studied by finite size scaling techniques.
\begin{itemize}
\item[I] In the region  of low  $\beta$,
at $N_s$  larger than any
physical length scale $\rho$ becomes $N_s$ independent and finite, and
from eq(63) one concludes that $\langle \mu\rangle \neq 0$ (fig.4).

\begin{figure}[htb]
\includegraphics[width = 0.8\linewidth, angle = 270]{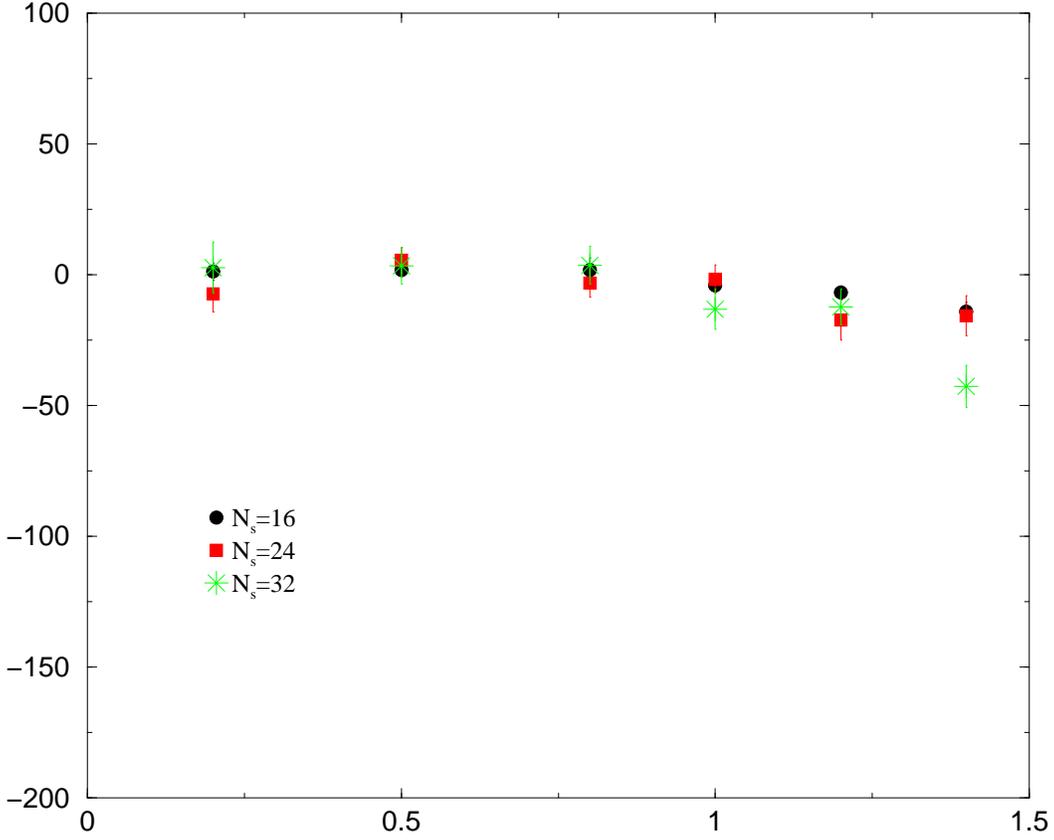}
\caption{Limit at large $N_s$ of $\rho$ for $\beta<\beta_c$, $SU(2)$ gauge group.}
\end{figure}

\item[II].In the region $\beta >\beta_c$  the dependence of $\rho$
   on $N_s$ can be studied . It is found (see fig.5)
 \begin{equation}
\rho  = - k N_s + k'\qquad k >0
\label{eq64}\end{equation}

\begin{figure}[htb]
\includegraphics[width = 0.8\linewidth]{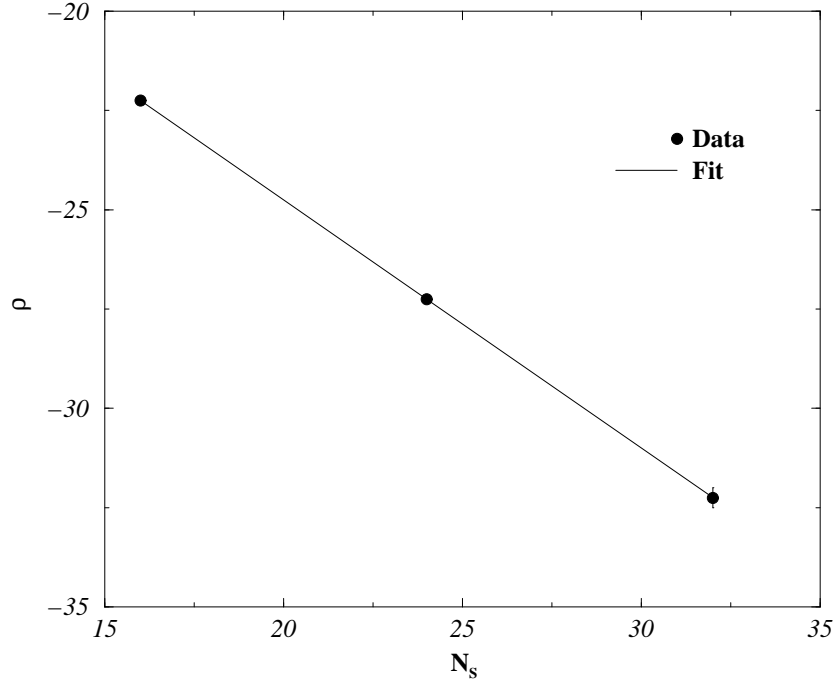}
\caption{$\rho$ vs $N_s$ for $\beta > \beta_c$.}
\end{figure}

In the limit $N_s\to\infty$,  $\langle\mu\rangle$ is strictly zero.  
 A direct determination of $\langle\mu\rangle$
 would give  zero   within large errors.
\item[III]
 In the vicinity of $\beta_c$  by 
dimensional arguments
 \begin{equation}
\langle\mu\rangle = \tau^\delta \Phi(\frac{a}{\xi},\frac{N_s}{\xi})
\label{eq65}\end{equation}
where $\xi$  is the correlation length in units of lattice spacings.
$\xi$  diverges at the critical point with an index $\nu$
 \begin{equation}
\xi \mathop\sim_{\tau\to 0}\tau^{-\nu}\qquad \tau = 1 -\frac{T}{T_c}
\label{eq66}\end{equation}
If $\xi\gg1$   then $ 1/\xi\sim0$     and
 \begin{equation}
\langle\mu\rangle \sim\tau^\delta\Phi(0,\frac{N_s}{\xi})
\xi \mathop\sim_{\tau\to 0}\tau^{-\nu}
\label{eq67}\end{equation}
One can trade $N_s/\xi$  with  $N_s^{1/\nu}\tau$  and
\begin{equation}
\rho = -\frac{\delta}{\tau} + N_s^{1/\nu} \Phi'(0,N_s^{1/\nu}\tau)
\label{eq68}\end{equation}
or
 \begin{equation}
\rho/N_s^{1/\nu} = -\frac{\delta}{\tau N_s^{1/\nu}} + 
\Phi'(0,N_s^{1/\nu}\tau)
\label{eq69}\end{equation}
\end{itemize}
A scaling law results: $\rho/N_s^{1/\nu}$  is a universal function 
of  
$N_s^{1/\nu}\tau$, independent of $N_s$.
Fig.6   shows a typical dependence of  $\rho$ on  $N_s$.
Fig.7 shows the quality of scaling for SU(2).  A best square fit to the 
data allows
to determine  $\beta_c,\nu,\delta$.

\begin{figure}[htb]
\includegraphics[width = 0.9\linewidth, angle = 270]{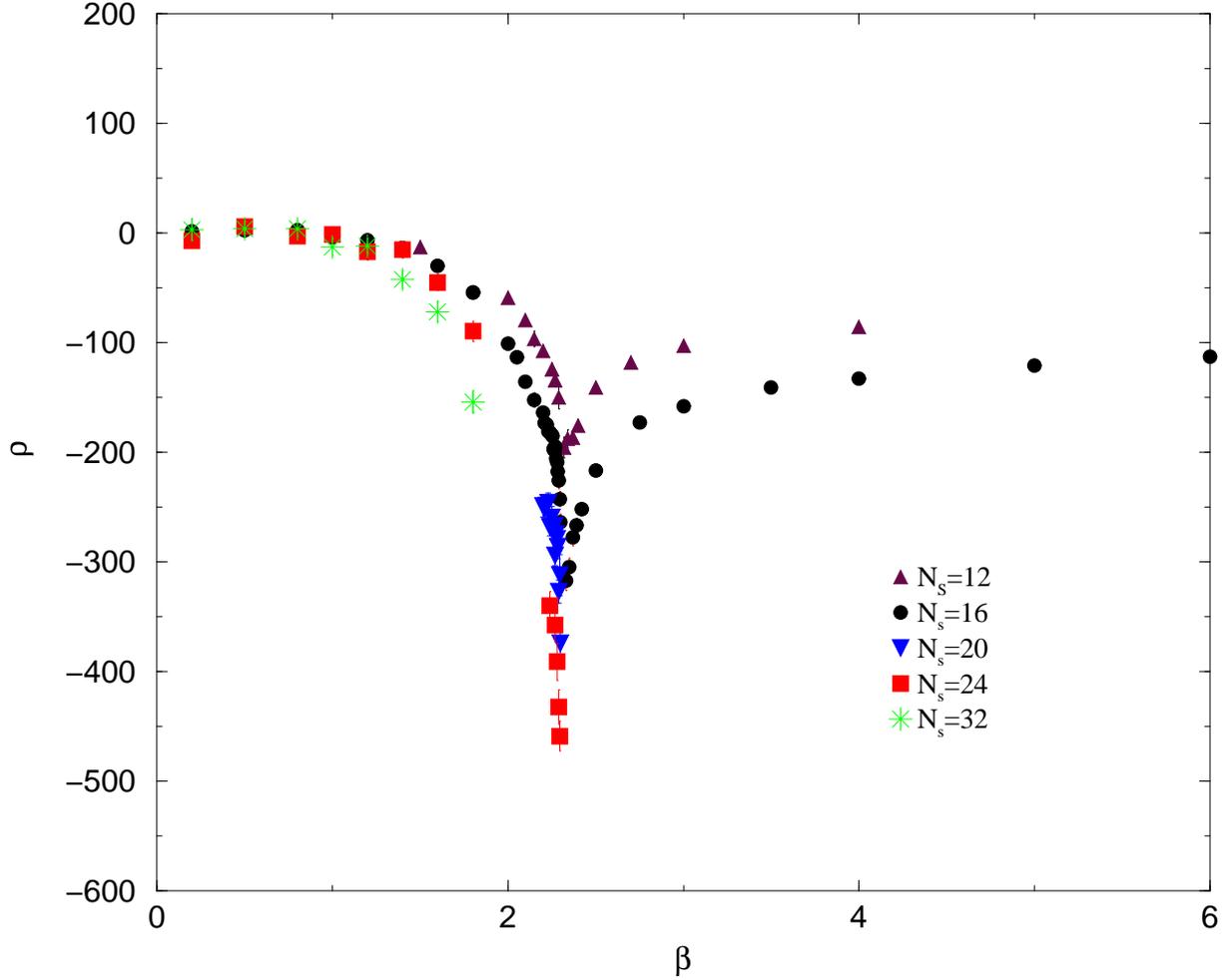}
\caption{Volume dependence of $\rho$, pure gauge $SU(2)$, deconfinement transition.}
\end{figure}

\begin{figure}[htb]
\includegraphics[width = 0.9\linewidth]{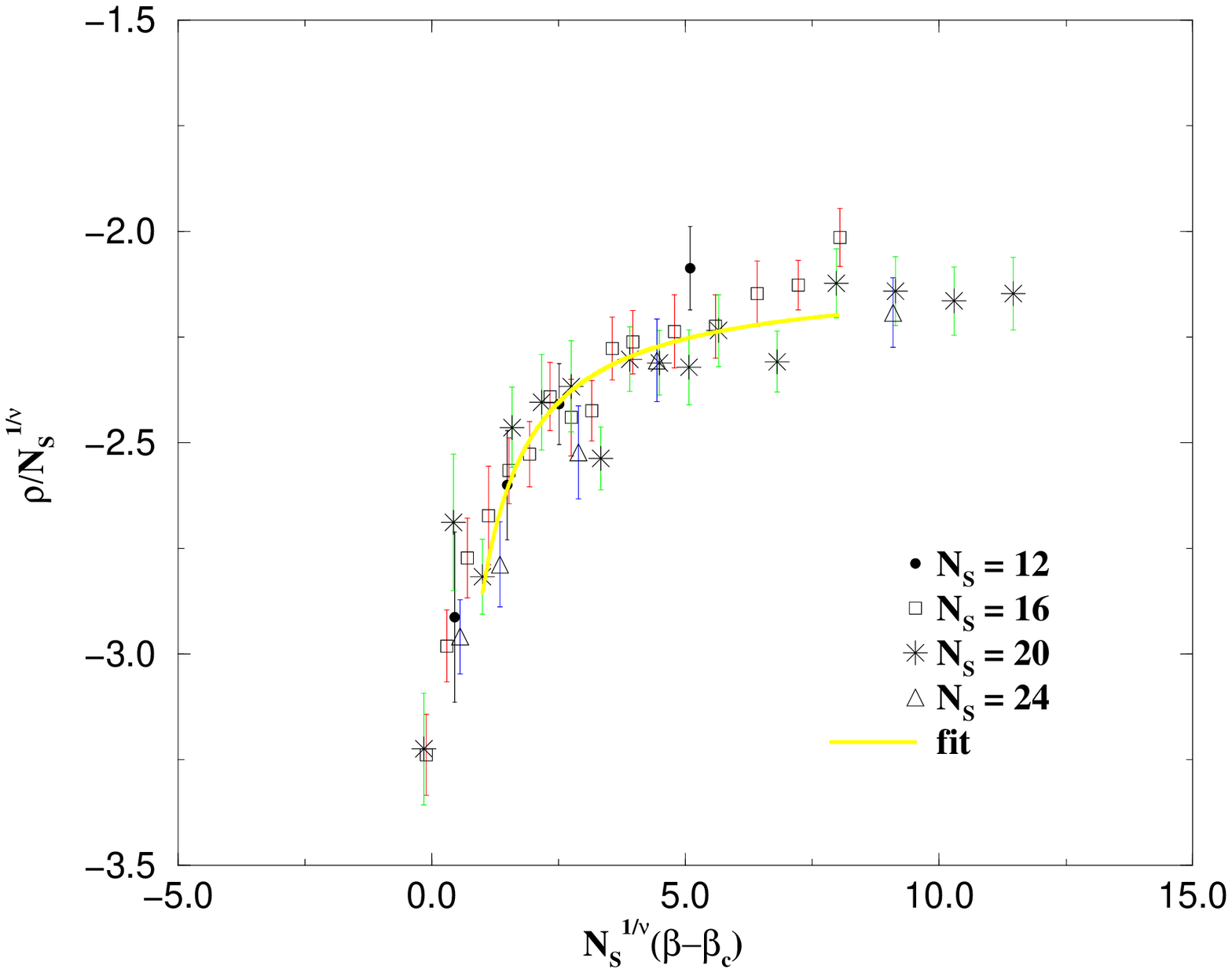}
\caption{Scaling, eq.(69). $SU(2)$ pure gauge.}
\end{figure}

For pure gauge theory one obtains\cite{35}
\[\begin{array}{lll}
SU(2)\hskip0.1in & \nu = .62(1) &\hskip0.1in \delta = 0.20(3)\\
SU(3) & \nu = .33(1) &\hskip0.1in \delta = .50(3)
\end{array}\]

The result is independent on the choice of the field   used to 
perform the abelian projection.
Confining vacuum is a dual superconductor in all abelian projections\cite{38}.

In all projections magnetic symmetry is Wigner in the deconfined 
phase.

Whatever the dual excitations of QCD are, they must be magnetically
charged in all the abelian projections.

In QCD with dynamical quarks the same operator 
$\langle\mu\rangle$
 can be defined ,and the corresponding  $\rho  = 
\frac{d}{d\beta}\log(\langle\mu\rangle)$
can be studied\cite{40}. Results already exist that  
$\langle\mu\rangle\neq0$
 in the confining phase, and  $\langle\mu\rangle=0$ in the infinite
volume limit.                      .
The finite size scaling in the vicinity of the transition is on the 
way.

Now another scale, the quark mass enters, which was absent in the 
quenched
case. Eq(65) -(67) become
\begin{equation}
\langle\mu\rangle \simeq \tau^\delta \Phi(\frac{1}{\xi},
\frac{\xi}{N_s}, m_q\xi) \mathop\simeq_{\xi\to\infty}
\tau^\delta \Phi(0,N-s^{1/\nu}\tau, m_q N_s^{1/\gamma})
\label{eq70}\end{equation}

\begin{figure}[htb]
\includegraphics[width = 0.9\linewidth]{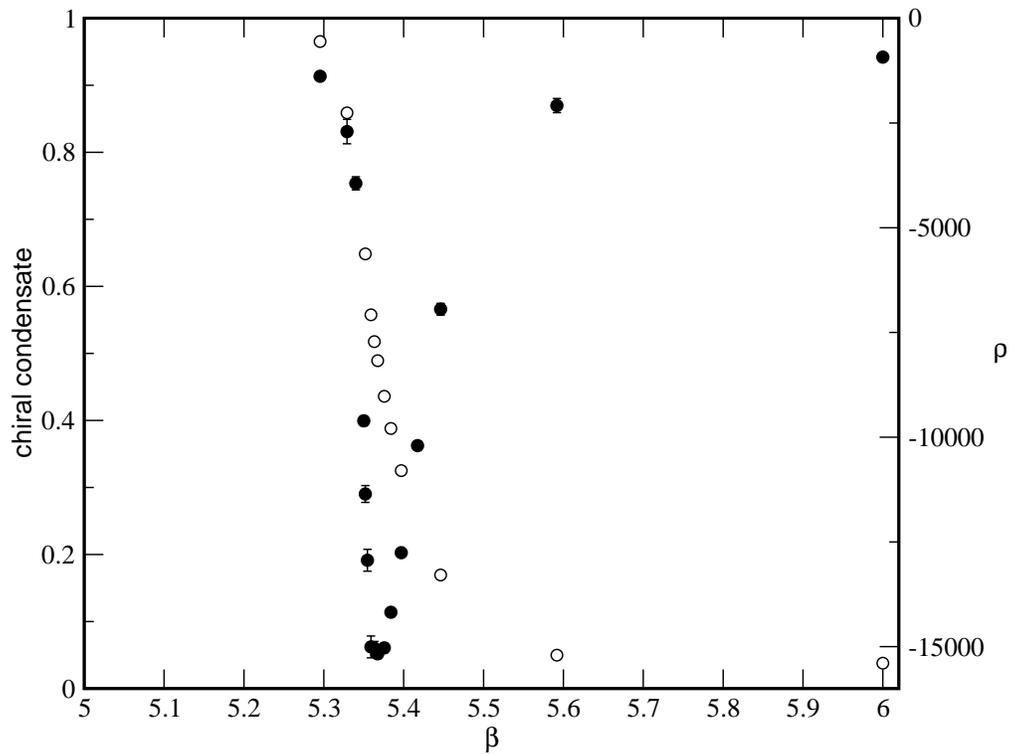}
\caption{2 flavour QCD. Open circles represent the chiral condensate, with
the scale axis on the left. Full circles represent $\rho$, with the scale 
on the right. The peak of $\rho$
coincides with the drop of $\langle\bar\psi\psi\rangle$.}
\end{figure}

The index $\gamma$ is known. In order to determine $\nu$  one can  choose 
values of  $N_s$  such that $m_q N_s^{1/\gamma}$=const.,
and test the scaling law
\begin{equation}
\rho/N_s^{1/\nu} \sim \Phi(N_s^{1/\nu}\tau)
\label{eq71}\end{equation}
to determine  $\nu$ and to investigate the nature of the transition. 
This will complete the analysis.
The existing data already  show,however,that dual superconductivity 
is the
mechanism of confinement also in the presence of dynamical quarks,
in agreement with the ideas of  $N_c\to\infty$. Fig.8 shows the negative peak of $\rho$
at the same temperature where the chiral condensate drops to zero\cite{40}.

\section{Concluding remarks.}
Large  distance  QCD  is an intrinsecally non perturbative 
system.Lattice provides
a tool to investigate this regime. 

Much progress has been done 
towards
the understanding of confinement. The mechanism of confinement is definitely dual 
superconductivity
both for pure gauge and full QCD. The dual excitations which condense 
to
produce confinement are not yet identified:what is known is that they 
are
magnetically charged in all abelian projections.

\begin{theacknowledgments}
I am indebted to   L Del Debbio, M.D'Elia ,
B.Lucini, E Meggiolaro, G.Paffuti, who have collaborated to the results 
presented
in these lectures. I am also grateful to Sergio  Lerma for help in preparing these notes
\end{theacknowledgments}

\end{document}